\documentclass[twocolumn]{aastex63}

\shorttitle{H$_2$ in AU Mic}
\shortauthors{Flagg et al.}

\begin{document}

\title{The Mysterious Affair of the H$_2$ in AU Mic}

\author[0000-0001-6362-0571]{Laura Flagg}
\affil{Department of Physics and Astronomy, Rice University, 6100 Main St. MS-108, Houston, TX 77005, USA}
\affil{Department of Astronomy and Carl Sagan Institute, Cornell University, Ithaca, New York 14853, USA}
\email{laura.flagg@cornell.edu}

\author[0000-0002-8828-6386]{Christopher M. Johns-Krull}
\affil{Department of Physics and Astronomy, Rice University, 6100 Main St. MS-108, Houston, TX 77005, USA}

\author[0000-0002-1002-3674]{Kevin France}
\affil{Laboratory for Atmospheric and Space Physics, University of Colorado, 600 UCB, Boulder, CO 80309, USA}

\author[0000-0002-7154-6065]{Gregory Herczeg}
\affil{Kavli Institute for Astronomy and Astrophysics, Peking University, Yi He Yuan Lu 5, Haidian Qu, Beijing 100871, China}

\author{Joan Najita}
\affil{NOIRLab, 950 Cherry Avenue, Tucson, AZ. 85719, USA}

\author[0000-0002-1176-3391]{Allison Youngblood}
\affil{Laboratory for Atmospheric and Space Physics, University of Colorado, 600 UCB, Boulder, CO 80309, USA}

\author[0000-0002-9540-853X]{Adolfo Carvalho}
\affil{Cahill Center for Astronomy and Astrophysics, California Institute of Technology, Pasadena, CA 91125, USA}

\author[0000-0003-2251-0602]{John Carptenter}
\affil{Joint ALMA Observatory, Avenida Alonso de $\rm{\acute{C}}$ordova 3170, Santiago, Chile}

\author[0000-0003-0214-609X]{Scott J. Kenyon}
\affil{Smithsonian Astrophysical Observatory, 60 Garden Street, Cambridge, MA 02138 USA}

\author[0000-0003-4150-841X]{Elisabeth Newton}
\affil{Department of Physics and Astronomy, Dartmouth College, Hanover, NH 03755, USA}

\author[0000-0003-1337-723X]{Keighley Rockcliffe}
\affil{Department of Physics and Astronomy, Dartmouth College, Hanover, NH 03755, USA}



\begin{abstract}
Molecular hydrogen is the most abundant molecule in the Galaxy and plays important roles for planets, their circumstellar environments, and many of their host stars.  We have confirmed the presence of molecular hydrogen in the AU Mic system using high-resolution FUV spectra from HST-STIS during both quiescence and a flare.  AU Mic is a $\sim$23 Myr M dwarf which hosts a debris disk and at least two planets. We estimate the temperature of the gas at 1000 to 2000 K, consistent with previous detections. Based on the radial velocities and widths of the H$_2$ line profiles and the response of the H$_2$ lines to a stellar flare, the H$_2$ line emission is likely produced in the star, rather than in the disk or the planet. However, the temperature of this gas is significantly below the temperature of the photosphere ($\sim$3650 K) and the predicted temperature of its star spots ($\gtrsim$2650 K). We discuss the possibility of  colder star spots or a cold layer in the photosphere of a pre-main sequence M dwarf.
\end{abstract}



\section{Introduction}\label{intro}

Planetary systems undergo dramatic changes for the first $\sim$100 Myr after their formation.  How a given planet evolves is a direct function of how both the host star and any circumstellar disk evolve and how they affect each other.  In order to study such complex interactions, observations of systems with circumstellar disks and planets are needed.  



One important issue is the state of the gas in the inner disk.  Because gas, especially warm gas, is hard to detect unless there are large amounts present, much less is known about the  evolution of gas in the inner disk once gas stops accreting onto the star  \citep[e.g.][]{HughesDebrisDisksStructure2018}. Observationally, it has been hard to distinguish between a reduction of the mass of gas and the complete absence of gas \citep{FlaggDetectionH2TWA2021}. Traditionally, it has been thought that the gas has completely dissipated once accretion is no longer detectable, but recent observations of systems like \object{TWA 4} \citep{YangFarultravioletAtlasLowresolution2012} or \object{TWA 7} \citep{FlaggDetectionH2TWA2021} show that this is not always true. 

One potential avenue to search for small amounts of H$_2$ is far-UV  observations \citep{InglebyFarUltravioletH2Emission2009, AlcalaHSTspectrareveal2019}. Molecular hydrogen --- which is the dominant component of protoplanetary disks --- only has allowed transitions in the UV. The FUV ($\sim$1000-2000 \AA) is particularly sensitive to warm gas (see Section \ref{sec_temp}), such as the gas that could still be present in the inner regions of developing solar systems.  However, with limited spatial resolution, it can be difficult if not impossible to distinguish small amounts of circumstellar H$_2$ from the H$_2$ that exists in stars.  M dwarfs exhibit H$_2$ emission, possibly from their photospheres \citep{KruczekH2FluorescenceDwarf2017};  warmer stars, like the Sun, have H$_2$ in starspots \citep{JordanEmissionlinesH21978}, which are similar in temperature to the photospheres of M dwarfs.  

A way around this problem is to observe systems with well known inclinations that are not face-on.  In these systems, if the signal-to-noise ratio of the spectrum is high enough to trace the H$_2$ line profile, the shape of the profile can help indicate the origin of the H$_2$ \citep{KruczekH2FluorescenceDwarf2017}.  If, for example, the line profile is much broader than typical line profiles for the star, then the H$_2$ probably originates in a circumstellar disk, orbiting the star at high velocities.  However, the opposite is not necessarily true, as circumstellar gas further out may produce a narrow profile.

Based on these criteria, an obvious target for the study of H$_2$ is \object{AU Mic}.  AU Mic  is a  M0Ve star \citep[see Table 1 for additional properties]{PecautIntrinsicColorsTemperatures2013} that is part of the $\sim$23 Myr  $\beta$ Pic Moving Group \citep{BarradoyNavascuesAgebetaPictoris1999, MamajekagePictorismoving2014, ShkolnikAllskyComovingRecovery2017}.  It has an edge-on debris disk discovered by \citet{KalasDiscoveryLargeDust2004}. The dust in the debris disk has since been observed and imaged in the optical, NIR, FIR, and sub-millimeter/millimeter \citep[e.g.][]{KristHubbleSpaceTelescope2005,GrahamSignaturePrimordialGrain2007,WilnerResolvedMillimeterEmission2012,MacGregorMillimeterEmissionStructure2013,MatthewsAUMicDebris2015,WangGeminiPlanetImager2015}. 

 \begin{deluxetable}{lrrl}
 \tabletypesize{\normalsize}
 \tablewidth{0pt}
 \tablecaption{Parameters of the AU Mic System}\label{params} 
 \tablehead{
 \colhead{Parameter} & \colhead{    \hspace{.25cm}   } & \colhead{Value} & \colhead{Citation}
  }
 \startdata
SpT & & M0Ve & \citet{PecautIntrinsicColorsTemperatures2013} \\
RV (km/s) & & -4.25$\pm$0.24 & \citet{SchneiderACRONYMIIIRadial2019} \\
$v$sin$i$ (km/s) & & 8.7$\pm$2.0 & \citet{Plavchanplanetdebrisdisk2020}\\
M$_*$ (M$_\odot$) & & 0.50$\pm$0.03 & \citet{Plavchanplanetdebrisdisk2020}\\
age (Myr) & & $\sim$23 Myr & \citet{MamajekagePictorismoving2014} \\
T$_{\rm{eff}}$ (K) & & 3642$\pm$22 & \citet{PecautIntrinsicColorsTemperatures2013} \\
\enddata
 \end{deluxetable}

However, these are all observations of the dust content of the disk. The characteristics of any gas in the disk remain uncertain.  Planet formation is greatly influenced by the gas, and not just because gas is an important component of planets themselves. Gas influences the motion of the small dust grains in the disk via gas drag \citep[e.g.][]{WeidenschillingAerodynamicssolidbodies1977, YoudinStreamingInstabilitiesProtoplanetary2005}, induces spirals and rings   \citep{LyraFormationsharpeccentric2013}, and can alter planet orbits \citep[e.g][]{GoldreichEccentricityEvolutionPlanets2003, BaruteauPlanetDiskInteractionsEarly2014}.  The presence of warm gas in a disk may also explain the discrepancy between the terrestrial planet population and the lack of detected IR flux from giant impacts that should be associated with the formation of these planets, because gas in the planet-forming region can ``sweep away'' dust \citep{KenyonRockyPlanetFormation2016}, as models indicate that terrestrial planet formation during that stage should produce detectable IR excess. 

Unfortunately,  circumstellar gas around AU Mic has been hard to detect and characterize.  \citet{LiuSubmillimeterSearchNearby2004} searched for --- but failed to find --- CO J=3-2 in its disk using the SCUBA bolometer array.  \citet{RobergeRapidDissipationPrimordial2005} placed upper limits on the H$_2$ in the disk using FUV observations from FUSE (R$\sim$20,000; 905-1187 \AA) and STIS (R$\sim$46,000; 1144 to 1710 \AA). \citet{FranceLowMassH2Component2007} detected H$_2$ from the system during quiescence, and concluded that due to its relatively low temperature between 800 and 2000 K, the H$_2$ is in the disk, not the star. \citet{KruczekH2FluorescenceDwarf2017} also detected H$_2$ during quiescence.  Further upper limits on the amount of atomic of H, He, and C were obtained from X-ray observations by \citet{SchneiderXrayingAUMicroscopii2010}.    \citet{DaleyMassStirringBodies2019} calculated an upper limit of 1.7$\times$10$^{-7}$ to 8.7$\times$10$^{-7}$ M$_\oplus$ of cold CO with excitation temperatures between 10 and 250 K based on ALMA data. Overall, the gas content has been elusive to quantify or characterize. Based on current measurements, the amount of gas in AU Mic's disk is clearly quite low, and there is a possibility that the H$_2$ detected does not lie in the disk \citep{KruczekH2FluorescenceDwarf2017}.


Even prior to the discovery of AU Mic's disk,  the star was well-known for its flares, which were first detected in the optical by  \citet{KunkelExistenceUpperLimit1970};   since then, the flares have since been studied in the EUV, X-ray, and radio \citep[e.g.][]{MonsignoriFossiTimeresolvedExtremeUltravioletSpectroscopic1996, SmithFlaresobservedXMMNewton2005, MacGregorPropertiesDwarfFlares2020}.  Recently, two young Neptunes, AU Mic b and c have been discovered in transit around the star, at distances of 0.06 and 0.11 AU, respectively \citep{Plavchanplanetdebrisdisk2020, MartioliNewconstraintsplanetary2021}.  Due to its relative proximity, the star and disk are comparatively well-studied, making AU Mic a prototype for young M dwarf planetary systems.  Understanding its inner disk gas content would provide constraints on the gas available for the planets to accrete and help us understand what is driving the dynamics at this point in the system's evolution.    





\section{Observations}
We used HST-STIS FUV-MAMA spectra of AU Mic from August 1998 \citep{PaganoHSTSTISEchelle2000} and July 2020  taken with the E140M grating with the 0.2\arcsec\ X 0.2\arcsec\  aperture taken in \textit{timetag} mode.  The spectra cover 1144 to 1710 \AA\ with resolving power R$\sim$46,000 depending on the grating order.  The observations are summarized in Table \ref{obs}.   AU Mic was observed for a total of 10105.74 s in 1998 (PID: 7556; PI: J. Linsky) and 15463.073 s in 2020 (PID: 15836; PI: E. Newton). Due to the decreasing sensitivity of the instrument with time, based on \citet{CarlbergUpdatedTimeDependent2017} we extrapolate that by 2020 the instrument had between 70\% and 85\% of the sensitivity it had in 1998, depending on the order.  

  \begin{figure*}
\centering
\includegraphics[width=0.98\textwidth]{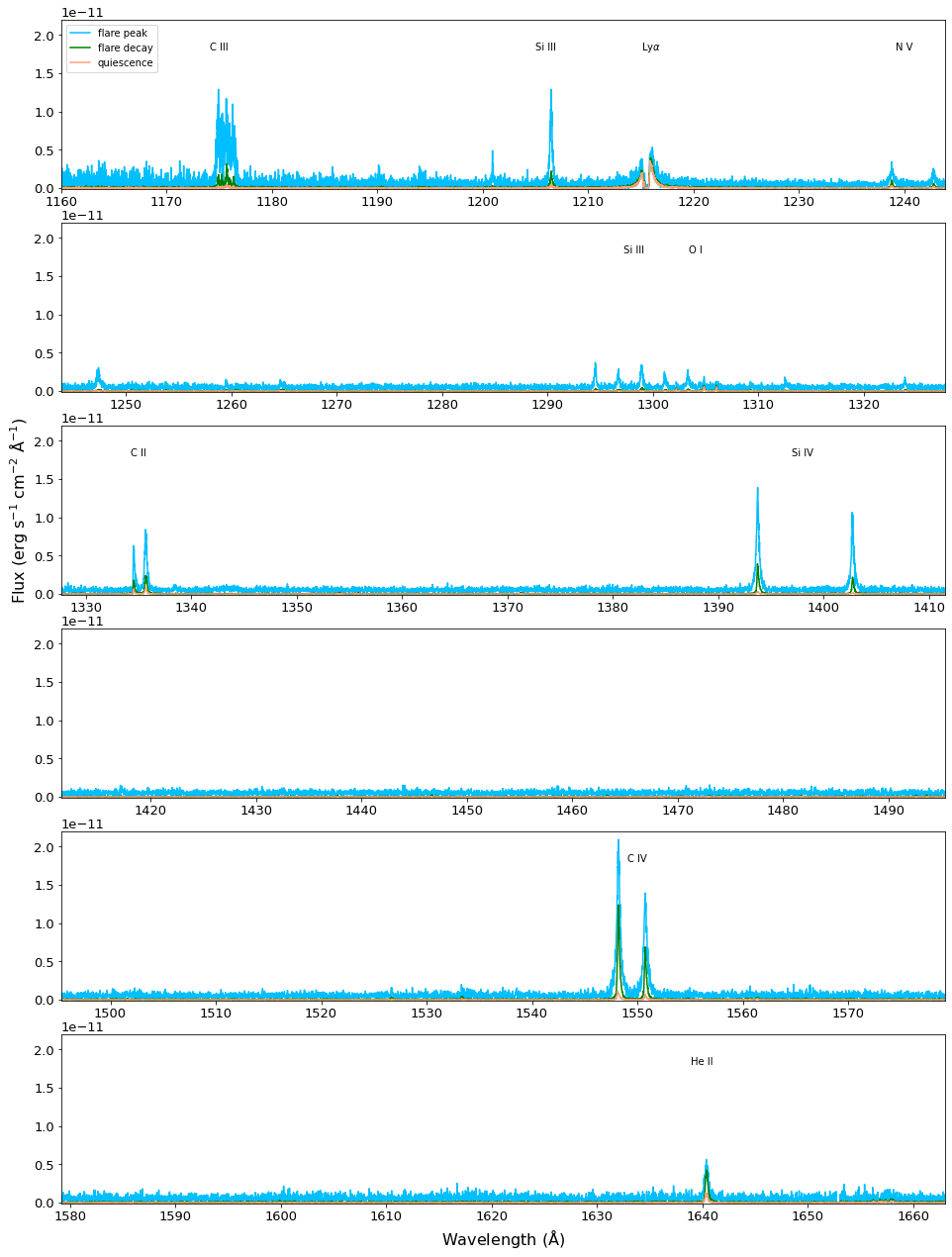}
\caption{The spectra from 2020: in quiescence compared to while flaring.  The features that change most visibly are hot chromospheric lines like Si IV or \ion{C}{4}. \label{f_v_q}}
\end{figure*}

\begin{table}[]
\centering
\begin{tabular}{llll}
   & \textbf{Exp Time} &    & \textbf{MJD mid}     \\
    \textbf{Dataset}      & (s)      &   \textbf{ID}     & (days)      \\
         \hline
         \hline
O4Z301010 & 2130.180  & 7556  & 51062.52430 \\
O4Z301020 & 2660.189 & 7556  & 51062.58529 \\
O4Z301030 & 2660.189 & 7556  & 51062.65248 \\
O4Z301040 & 2655.182 & 7556  & 51062.71963 \\
\hline
OE4H01010 & 2306.173 & 15836 & 59032.03795 \\
OE4H01020 & 2848.183 & 15836 & 59032.10325 \\
OE4H02010 & 2306.188 & 15836 & 59032.24029 \\
OE4H02020 & 2848.192 & 15836 & 59032.30198 \\
OE4H02030 & 2848.169 & 15836 & 59032.36822 \\
OE4H03010 & 2306.168 & 15836 & 59033.03164
\end{tabular}
\caption{Observation of AU Mic taken with HST-STIS using the E140M grating with the 0.2\arcsec\ X 0.2\arcsec\ aperture.  The horizontal line divides the data taken in August 1998 (top) from the data taken in July 2020 (bottom). \label{obs}}
\end{table}

The spectra were reduced with the STIS pipeline.\footnote{https://github.com/spacetelescope/stistools}  We then interpolated each observation onto a common wavelength scale.  We then did an initial analysis of the observations from each HST orbit separately.  During the 2020 observations, there was a significant flare during the first exposure (Figure \ref{f_v_q}), and two of the later exposures were taken during a transit of AU Mic b.  We therefore analyzed those exposures separately; the other three exposures were coadded for further analysis.  There was also a more minor flare during the first exposure of the 1998 data, analyzed by  \citet{RobinsonFarUltravioletObservationsFlares2001} and smaller flares that were still  noticeable by eye during the second exposure. We coadded only the data from the two remaining exposure in 1998 for the analysis of the temperature (Section \ref{sec_temp}) as the temperature would be specially sensitive to the flare; all the data from 1998 was coadded for the purpose of analyzing the line profile, as presented in Sections \ref{subsubsec_lsddisk} and \ref{subsec_star}.

\section{\texorpdfstring{H$_2$ Detection and Verification in Quiescence}{H2 Detection and Verification in Quiescence}}\label{sec:h2detect}
We used the FUV spectra to detect H$_2$ during quiescence.  The spectra are dominated by chromospheric lines, such as the ones noted in Figure \ref{f_v_q}.   During quiescence, the individual   H$_2$ features are buried in the continuum noise and are not bright enough to be detected on their own.  Instead, we used two methods to combine the signals from multiple features: least-squares deconvolution (LSD), as implemented by \citet{ChenSpectropolarimetryClassicalTauri2013}, and a cross-correlation function (CCF).  LSD is a way of extracting the average shape of the line profile from many lines across a spectrum \citep{DonatiSpectropolarimetricobservationsactive1997}.  Both methods require a selection of  H$_2$ lines \citep{AbgrallTableLymanBand1993} and their expected line strengths, for which we used models from \citet{McJunkinEmpiricallyEstimatedFarUV2016}.   We also set a minimum peak line intensity for each method to maximize the signal-to-noise of our result.  Using too many weak lines in both the CCF and the LSD will increase the noise more than the signal.  The specific minimum peak line intensity depends both on the method and the set of H$_2$ lines.  LSD profiles and CCFs are sensitive to noise in different manners, so we chose a  slightly different minimum peak line intensity based on what was appropriate for each method.  We also looked at individual progressions, which are H$_2$ emission lines from the same excited state [\textit{v'},\textit{J'}],\footnote{We use the notation [\textit{v'},\textit{J'}] to describe a progression, where \textit{v'} and \textit{J'} are the vibrational and rotational levels respectively in the first excited electronic state for a given progression.} thus changing the set of H$_2$ lines. 

\begin{figure*}
\centering
\includegraphics[width=0.98\textwidth]{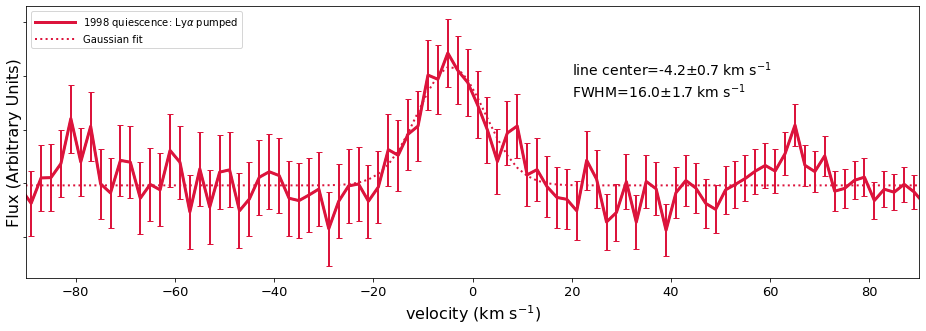}
\caption{The line profile reconstructed using LSD fit with a Gaussian. The line center is consistent with the systemic velocity of the star. \label{profile_1998}}
\end{figure*}

\begin{figure*}
\centering
\includegraphics[width=0.98\textwidth]{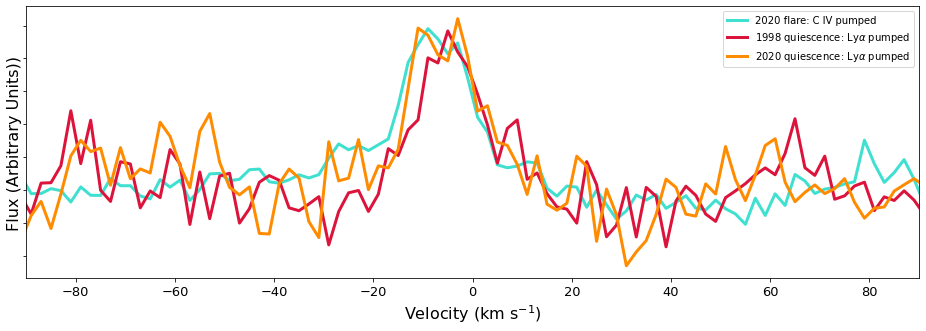}
\caption{Comparison between the reconstructed line profiles for 1998 during quiescence, 2020 during quiescence, and 2020 during the flare.  Note: the profiles are scaled to similar heights. \label{lsd}}
\end{figure*}

For the LSD, (unlike the CCF described below) we are able to extract line profiles and associated uncertainties directly, using a minimum peak line intensity  of 1$\times$10$^{-16}$ erg s$^{-1}$ cm$^{-2}$ \AA$^{-1}$, resulting in one LSD profile combining all progressions for the 1998 data and another profile for the 2020 data. We can then do a basic analysis of these profiles by fitting Gaussians to them, as shown in Figure \ref{profile_1998} for the data from 1998. The standard errors on the Gaussian fit are calculated from the covariance matrix.  The line center from the LSD profile, -4.2$\pm$0.7 km/s, is consistent within uncertainties with the systemic velocity of AU Mic of -4.25$\pm$0.24 km/s \citep{SchneiderACRONYMIIIRadial2019}.  The FWHM of the line is 16.0$\pm$1.7 km/s.   Figure \ref{lsd} shows the resulting LSD profiles from 2020 compared to that from 1998.  We see relatively similar profiles for the 1998 and 2020 data, as well as the 2020 flare, although the line profile from the 2020 data is slightly blue shifted. The Gaussian fits to all three profiles are summarized in Table \ref{profile_pars}.

\begin{table}[]
\centering
\begin{tabular}{lll}
   \textbf{} &  \textbf{Center}    & \textbf{FWHM}   \\
    \textbf{Profile}      & (km s$^{-1}$)      &   (km s$^{-1}$)      \\
         \hline
1998 Quiescence & -4.2$\pm$0.7 & 16.0$\pm$1.7  \\
2020 Quiescence & -5.7$\pm$0.8 & 16.6$\pm$1.9   \\
2020 Flare & -7.7$\pm$0.5 & 19.0$\pm$1.2   \\
\end{tabular}
\caption{Fit parameters and uncertainties for all three profiles.\label{profile_pars}}
\end{table}

\begin{figure*}
\centering
\includegraphics[width=0.98\textwidth]{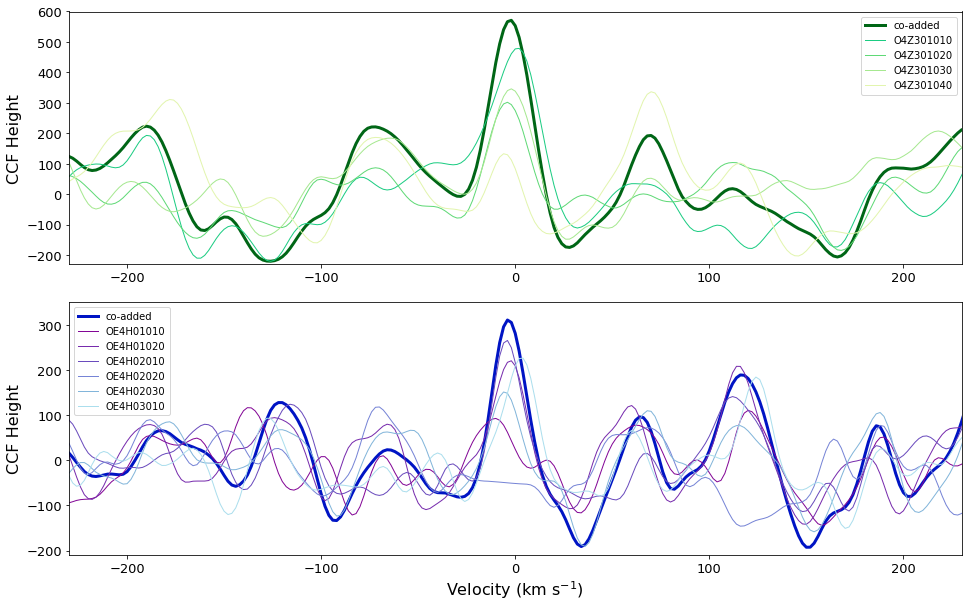}
\caption{The CCFs from 1998 (top) and 2020 (bottom) both show a clear detection of H$_2$ from  Ly$\alpha$ pumped progressions.  The thin lines are each separate orbit, with labels that correspond to the individual observations listed in Table \ref{obs}.  The thick lines are the CCF from the co-added spectrum. \label{lya_ccf}}
\end{figure*}

Our procedure for creating the CCF of AU Mic is based on  \citet{FlaggDetectionH2TWA2021}.  To summarize: we masked out the hot gas lines from the star and then cross-correlated the masked stellar spectrum with H$_2$ templates. For this analysis, we used what Flagg et al. refers to as a segmented spectrum, a spectrum with only segments that contain H$_2$ features that are expected to be prominent. The segmented spectrum preserves the relative line heights of different H$_2$ features, which are needed to measure a temperature (see Section \ref{sec_temp}).  In the data from 1998 and 2020, we clearly detect four progressions pumped by Ly$\alpha$: [1,4]  --- detected previously by \citet{KruczekH2FluorescenceDwarf2017} --- as well as  [1,7], [0,1], and [0,2], as shown in Figure \ref{lya_ccf}.

\section{\texorpdfstring{H$_2$ Detection During a Stellar Flare}{H2 Detection During a Stellar Flare}}
\begin{figure*}
\centering
\includegraphics[width=0.98\textwidth]{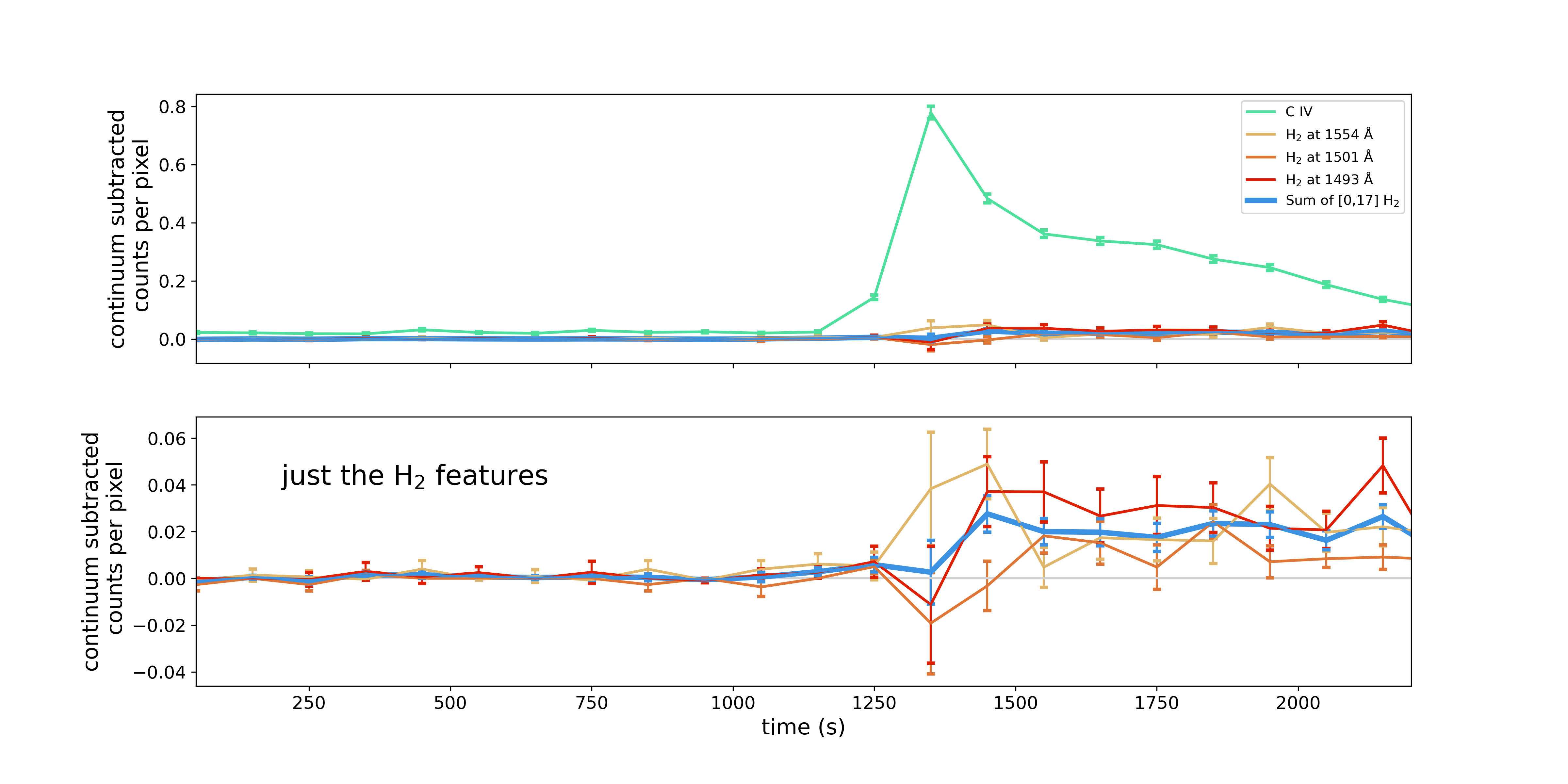}
\caption{Continuum subtracted light curves for different spectral features for observation OE4H01010 acquired on 2020 July 2.  \ion{C}{4} (light green line, top plot), which is a hot transition region line, peaks at 1350 s.  The various H$_2$ features (shown more clearly in the bottom plot), are all in the [0,17] progression and pumped by \ion{C}{4}.  The individual features have all clearly flared by 1550 s at the latest, with one feature flaring as early as 1350 s, and their combined brightness flaring by 1450 s, implying at most a 200 s delay between the \ion{C}{4} increasing in brightness and the resulting increase in brightness from the H$_2$.   \label{lc_flare}}
\end{figure*}

\begin{figure*}
\centering
\includegraphics[width=0.98\textwidth]{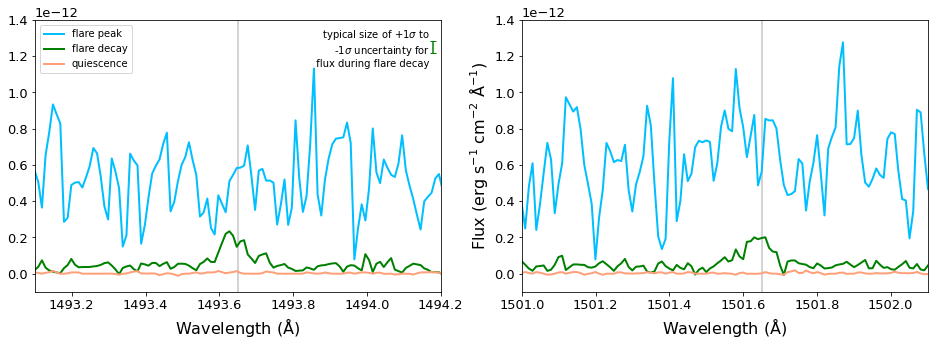}
\caption{Spectra from 2020.  The observation during the  flare decay (green line) show emission from H$_2$ features not seen during quiescence (orange line). During the flare peak, the noise in the continuum makes detecting any H$_2$ difficult.   \label{h2_f_v_q}}
\end{figure*}

The response of the H$_2$ emission during a stellar flare gives insight as to the nature and source of the H$_2$.  For both 1998 and 2020, the first exposure contains a stellar flare.   In 1998, the flare  \citep[analyzed in detail by][]{RobinsonFarUltravioletObservationsFlares2001}, was fairly weak, with flux from hot chromospheric lines like \ion{C}{4} increasing by less than a factor of 2.  In comparison, the 2020 flare data in observation OE4H01010 was much stronger, with fluxes in chromospheric lines increasing by a factor of $\sim$40, as shown in Figure \ref{lc_flare}.  During both flares,  we detect H$_2$ emission in the spectrum that is not detectable during quiescence. As the 2020 flare was significantly brighter than the 1998 flare, the H$_2$ was correspondingly brighter, so we focused our analysis on the 2020 data set. We show the spectra of two prominent H$_2$ features that flared in the 2020 data, both during the flare and in quiescence, in Figure  \ref{h2_f_v_q}.

\begin{figure*}
\centering
\includegraphics[width=0.98\textwidth]{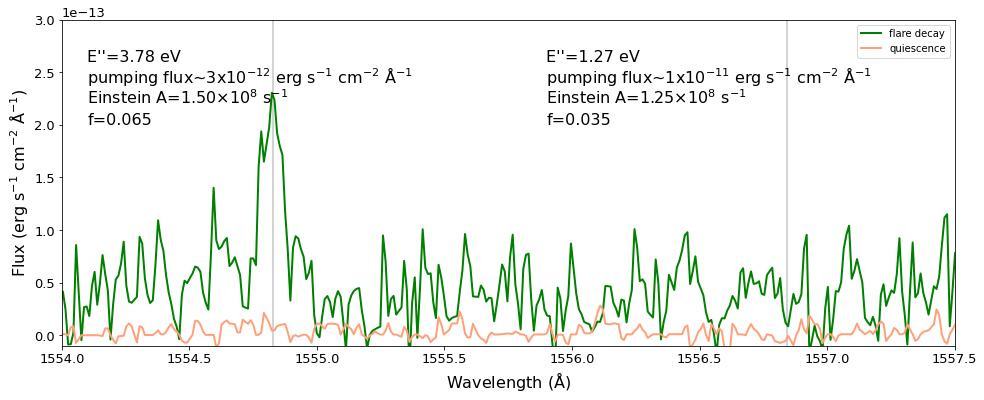}
\caption{Two prominent H$_2$ features from [0,17] at 1554.8 \AA\ and [1,7] at 1556.9 \AA\, marked by vertical gray lines.  During the flare decay (green line, observation OE4H01010 acquired on 2020 July 2), only the one that originates from a much high energy level is detectable by eye, implying the  H$_2$ level populations are affected by a  non-thermal mechanism during the flare. \label{sbs}}
\end{figure*}

The H$_2$ features  detectable by eye during the 2020 flare are from the [0,17] or the [0,24] progressions. (The spectra from all features are shown in the appendix.) They are pumped by the \ion{C}{4} doublet at 1550 \AA, whose brightness increased by $\sim$40x during the flare (Figure \ref{f_v_q}). The detection of the resulting H$_2$ lines is surprising, although not unprecedented \citep{HerczegOriginsFluorescentH22006}, because these H$_2$ lines originate from states with energies of 3.78 eV and 4.19 eV above ground, compared to between 1.0 and 1.3 eV for lines in progressions we detect during quiescence. For example, during the flare, the H$_2$ feature at 1554.8 \AA\ is quite strong, while the H$_2$ feature at 1556.9 \AA\ is not  (Figure \ref{sbs}).  In thermal equilibrium at temperatures below 10000 K, these flux ratios are not possible, because the line at 1556.9 \AA\ is populated from a state with energy of 1.27 eV from the [1,7] progression while the line at 1554.8 \AA\ is from the [0,17] progression populated from a state at 3.78 eV.    Clearly, the flare results in non-thermal populations of H$_2$. 

Overall, we detect flux from eight features from [0,17] and two features from [0,24], which are summarized in Table \ref{civ_detections}.  These were the only H$_2$ progressions that clearly had extra emission during the flare. The flux from Ly$\alpha$ did not increase substantially during the flare (Figure \ref{f_v_q}), consistent with the findings from \citet{LoydMUSCLESTreasurySurvey2018}, and thus the H$_2$ lines that are pumped by Ly$\alpha$ show no significant increase in flux. 


 \begin{deluxetable*}{cccrrcr}
 \tabletypesize{\normalsize}
 \tablewidth{0pt}
 \tablecaption{\ion{C}{4} Pumped H$_2$ Detections During Flare}\label{civ_detections} 
 \tablehead{
 \colhead{Wavelength} & \colhead{     } & \colhead{lower E} & \colhead{Flux$\times$10$^{-15}$}& \colhead{Einstein A} & \colhead{Velocity} & \colhead{FWHM}
 \vspace{-5pt}
 \\
  \colhead{(\AA)} & \colhead{  Progression   } & \colhead{(eV)} & \colhead{(erg s$^{-1}$ cm$^{-2}$)}& \colhead{(s$^{-1}$)} & \colhead{(km s$^{-1}$)} & \colhead{(km s$^{-1}$)}
  }
 \startdata
1501.67	&	[0,17]	&	3.78	&	16.1$\pm$	2.8	&	21.78	&	-7.6	&	18.6	\\
1493.67	&	[0,17]	&	3.78	&	17.1$\pm$	4.7	&	21.07	&	-7.1	&	13.1	\\
1446.72	&	[0,17]	&	3.78	&	14.6$\pm$	3.6	&	19.48	&	-9.1	&	17.1	\\
1554.85	&	[0,17]	&	3.78	&	20.1$\pm$	7.1	&	14.97	&	-6.1	&	14.1	\\
1437.78	&	[0,17]	&	3.78	&	11.7$\pm$	2.8	&	14.35	&	-9.3	&	17.8	\\
1391.01	&	[0,17]	&	3.78	&	7.8$\pm$	2.0	&	11.32	&	-8.4	&	12.8	\\
1599.93	&	[0,17]	&	3.78	&	20.2$\pm$	10.6	&	9.59	&	-4.9	&	11.2	\\
1381.41	&	[0,17]	&	3.78	&	3.3$\pm$	2.5	&	6.10	&	-5.2	&	14.4	\\
\hline
1594.05	&	[0,24]	&	4.19	&	4.2$\pm$	3.6	&	23.06	&	-8.8	&	7.3	\\
1586.69	&	[0,24]	&	4.19	&	7.7$\pm$	4.5	&	21.54	&	-4.0	&	17.2	
\enddata
\tablecomments{The listed velocity is relative to the listed wavelengths, which may be inaccurate by a few km/s.}
 \end{deluxetable*}

\section{\texorpdfstring{Temperature of H$_2$ in Quiescence}{Temperature of H2 in Quiescence}}\label{sec_temp}
  The temperature of the H$_2$ emission helps to constrain its origin. We estimate the gas temperature by analyzing the H$_2$ emission, assuming the gas is thermally populated while in quiescence. However, we cannot derive a gas temperature directly from H$_2$.  Whether or not we detect flux from a progression and how much flux we see depends on several factors:
\begin{enumerate}
        \item a populated lower state of the pumping transition of H$_2$ molecules 
    \item a pumping transition with a relatively high oscillator strength
    \item flux to excite the H$_2$ molecule into the higher states
    \item the Einstein A values of the decaying transitions for a progression
    \item the filling factor of the gas
\end{enumerate}
Items 2) and 4) are solely dependent on molecular physics and therefore are well known. Item 3) depends on our knowledge of the flux at the pumping wavelength.  In the case of H$_2$ lines pumped by hot chromospheric lines, this is typically trivial, because we directly observe those pumping lines. However, the Ly$\alpha$ line profile is contaminated by ISM absorption, so for transitions pumped by Ly$\alpha$,  we need to reconstruct the Ly$\alpha$ line profile in order to estimate this flux.  This was carried out for AU Mic following the methods of \citet{YoungbloodFUMESIILya2021}. We derive a profile with an intrinsic integrated flux of 8.94$\times$ 10$^{-12}$ erg s$^{-1}$ cm$^{-2}$, in agreement with the value reported in \citet{YoungbloodMUSCLESTreasurySurvey2016}. The difference between the two reconstructions is the approximation of the intrinsic emission line by a Voigt profile, which matches the broad wings better than the double Gaussian approach of \citet{YoungbloodMUSCLESTreasurySurvey2016}. Item 5) is unknown, but is often assumed to be 1 \citep{McJunkinEmpiricallyEstimatedFarUV2016}. This leaves only item 1).   If the H$_2$ lines are optically thin, which is the case for the low levels of emission we detect from AU Mic,  and  the H$_2$ is thermally populated in quiescence --- which is consistent with models \citep{AdamkovicsFUVIrradiatedDisk2016} --- then the relative fluxes of H$_2$ features directly trace the excitation temperature of the gas.  An estimation of the temperature of the H$_2$ could help constrain its location.

\begin{figure*}
\centering
\includegraphics[width=0.98\textwidth]{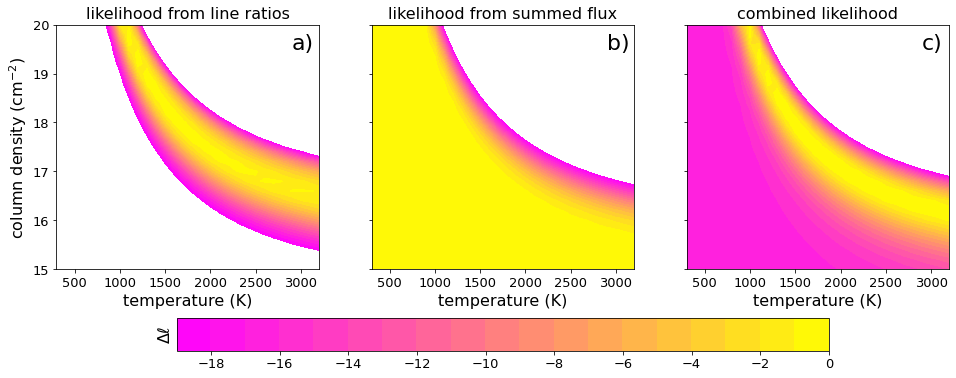}
\caption{Log likelihoods for the temperature and H$_2$ column density of the gas.  \label{ll}}
\end{figure*}
\begin{figure*}
\centering
\includegraphics[width=0.98\textwidth]{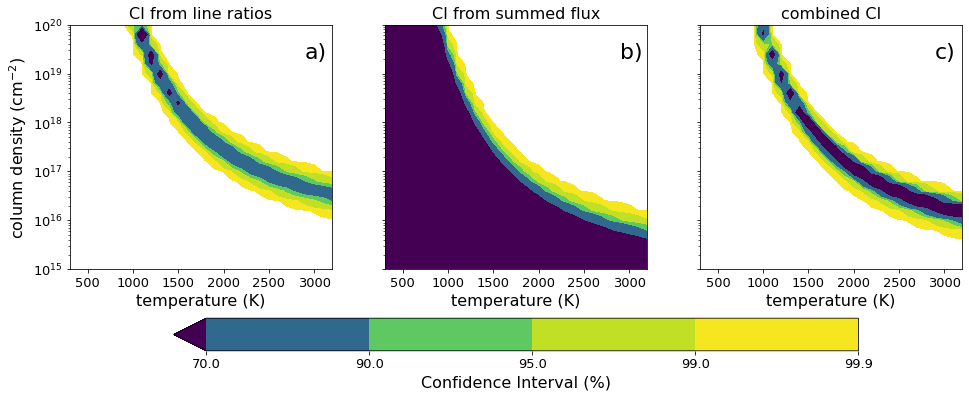}
\caption{Confidence intervals for the temperature and H$_2$ column density of the gas.  \label{ci}}
\end{figure*}

We created spectral templates of H$_2$ fluorescence with AU Mic's Ly$\alpha$ profile as we did in Section \ref{sec:h2detect} using the models from \citet{McJunkinEmpiricallyEstimatedFarUV2016} with a uniform filling factor. Our grid of models covers from 300 K to 3200 K in 100 K increments at column densities between log($N$)=15  and log($N$)=20  in increments of 0.2 dex.  Because we are assuming the gas is thermally populated, we know the gas is at least 300 K and likely much warmer because of the ground state energy levels of the FUV H$_2$ transitions. For each template, we calculated the likelihood, L, (Figure \ref{ll}a) based on the CCF between the spectrum and each template as in \citet{BrogiRetrievingTemperaturesAbundances2019} with:
  \begin{equation}\label{bl_ll}
  \ell=  \log(L) = -\frac{N}{2}\log\left[ s^2 -2R + s_m^2 \right]-N\log(2\pi )
\end{equation}
where $N$ is the number of points in the spectrum, $s^2$ is the variance in the spectrum, $s_m^2$ is the variance in the model, and  $R$ is the cross-covariance. The cross-correlation function height is equal to $\frac{R}{\sqrt{s_f^2s_g^2}}$. We then translate that into confidence intervals based on the likelihood ratio test (Figure \ref{ci}a).  At these temperatures, the line ratios --- all that the CCF is sensitive to --- vary little, and the resulting uncertainties are large.

To complement this analysis, we also measure  total H$_2$ emission in the spectrum to better constrain the H$_2$ temperature. We co-added the flux in the regions of the strongest H$_2$ lines to result in a co-added line profile, integrated the flux from that profile, and fit that integrated flux to what we obtain from the same procedure (co-adding the strongest lines and integrating the flux of the resulting profile) for each model in the grid.  While the co-added flux is more sensitive to different column densities and temperatures, there are also much larger uncertainties in calculating the co-added flux. For example, uncertainties in the Ly$\alpha$ reconstruction result in only a few percent uncertainties in the ratios of amount of flux at the pumping wavelengths, but up to 30\% uncertainties in the absolute flux. (Uncertainties in Ly$\alpha$ reconstruction are dominated by the column density of hydrogen in the ISM, which generally increases or decreases \textit{all} the Ly$\alpha$ flux, thus having less effect on relative fluxes and more effect on absolute fluxes.) We also considered uncertainties from the continuum subtraction ($\sim$6.2$\times$ 10$^{-16}$ erg s$^{-1}$ cm$^{-2}$)  and  the flux uncertainties returned by the pipeline reduction ($\sim$3.8$\times$ 10$^{-15}$ erg s$^{-1}$ cm$^{-2}$). The final uncertainty in the flux measurement, $\sigma$, is the sum of all of these added in quadrature. The log-likelihood, $\ell$ is then calculated as:
  \begin{equation}\label{modfit_ll}
      \ell=  \log(L) =-\frac{1}{2}\left[ \frac{(F_{mod}-F_{meas})^2}{\sigma^2} +\log(2\pi \sigma^2)\right]
\end{equation}
where $F_{meas}$ is the flux measured from co-adding H$_2$ features in the data while $F_{mod}$ is the flux from co-adding the features in the model. This results in  log likelihoods for our models as shown in Figure \ref{ll}b with the corresponding confidence intervals in Figure \ref{ci}b.

Finally, we added the likelihoods from both the CCF and the co-added flux to produce the log likelihoods in Figure \ref{ll}c and the confidence intervals in Figure \ref{ci}c. Our best fit model has a log($N$)=17.6  and T=1900 K, consistent with the temperature range from \citet{FranceLowMassH2Component2007} of 800 to 2000 K;  our best column density estimate falls just outside the range from \citet{FranceLowMassH2Component2007} of 2.8$\times$10$^{15}$ to 1.9$\times$10$^{17}$ cm$^{-2}$.   Our 95\% confidence interval stretches beyond  the limits of our grid, but we conclude the gas is at $>$1000 K with a 99.9\% confidence. \citet{FranceLowMassH2Component2007} put an upper limit on the H$_2$ temperature based on the relative strength of O VI pumped lines to Ly$\alpha$ pumped lines:  only at temperatures below 2000 K do the O VI pumped lines dominate in the way they do in the FUSE spectrum. Thus  we adopt a temperature range of 1000 K to 2000 K for the H$_2$ during quiescence. 
  
  \section{Discussion}
  We consider four different possible origins for the detected H$_2$ emission: an unrelated background/foreground source, the disk, the planet, or the star.

      
  \subsection{Unrelated Source}
  An unrelated source, such as a background object or interstellar gas, would not receive any detectable amount of heating from a flare.  Thus, given that the H$_2$ flux increases as a response to the stellar flare, we can rule out a foreground or background source.  Even without the flare, a background source is unlikely.   Our detection is at the systemic RV of the AU Mic system, so it would have to be a source not only at the same RA and Dec, but also moving with the same velocity. 
  
  \subsection{Disk}
  While other gas species have been detected in circumstellar disks as old as AU Mic, H$_2$ has not been confirmed in any disks older than 15 Myr \citep[see review by][]{HughesDebrisDisksStructure2018}.  However, as mentioned in Section \ref{intro}, H$_2$ is hard to detect due to its homonuclear nature, so this could merely be an observational bias.  AU Mic is relatively nearby, which aids the ability to detect weak emission of H$_2$ in it. 
  
  
  \subsubsection{Analysis Based on the LSD Profile}\label{subsubsec_lsddisk}
  
  Line profiles from disks trace the location of the gas, because the gas velocity approximately follows Kepler's laws, so the velocity decreases with increasing distance from the star.  The average radius of the gas can be estimated by equation 2 from \citet{SchindhelmCharacterizingCOFourth2012}:
  \begin{equation}\label{ravg}
    r=GM_*\left(\frac{2sin(i)}{FWHM}\right)^2
\end{equation}
 where $r$ is the distance of the gas from the star, $M_*$ is the mass of the star, $G$ is the gravitational constant, and $i$ is the inclination of the disk. For AU Mic, the line profile from the H$_2$ during quiescence has a FWHM of 16.0$\pm$1.7 km/s, as discussed in Section \ref{sec:h2detect}, which  gives an average radius of $\sim$7 AU. This radius already poses problems, because given the flux from the star, heating the disk to T$>$1000 K at 7 AU  is impossible based on our current understanding of disk physics.

\begin{figure*}[h!]
\centering
\includegraphics[width=0.98\textwidth]{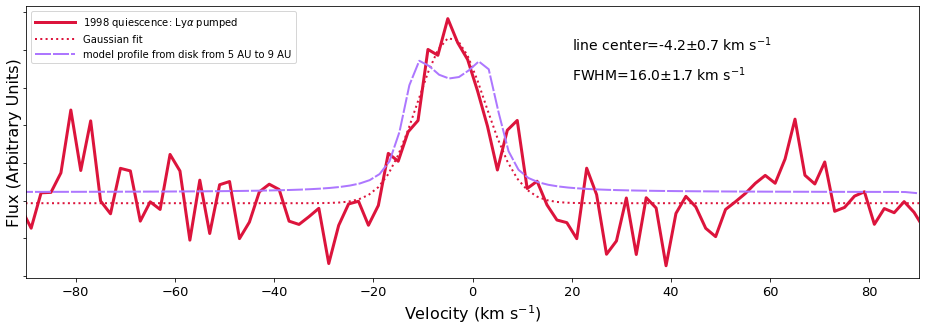}
\caption{The line profile reconstructed using LSD compared with a model disk profile, assuming the gas was in a ring from 5 to 9 AU. \label{lsd_disk}}
\end{figure*}

 We created a model profile  based on this average radius, using a flat, optically thick disk model between 5 and 9 AU, assuming a stellar mass of 0.5 M$_\odot$ \citep{Plavchanplanetdebrisdisk2020}.  Because we assume a flat disk, we also assume that the stellar flux absorbed --- and thus the corresponding H$_2$ flux ---  scales as $r^{-3}$.  This profile was then convolved with the line spread function (LSF) of the spectrograph. This modeled profile fits the LSD profile (Figure $\ref{lsd_disk}$) reasonably well with regards to the width.  However, the LSD profile lacks the  the characteristic ``M'' (double-peaked) shape profile from the line, which could indicate that the gas does not originate in the disk.    While gas at larger radii could result in a more Gaussian-like profile, that would make heating the gas even more difficult than it is around 7 AU. 
  
    \begin{figure}
\centering
\includegraphics[width=3.2in]{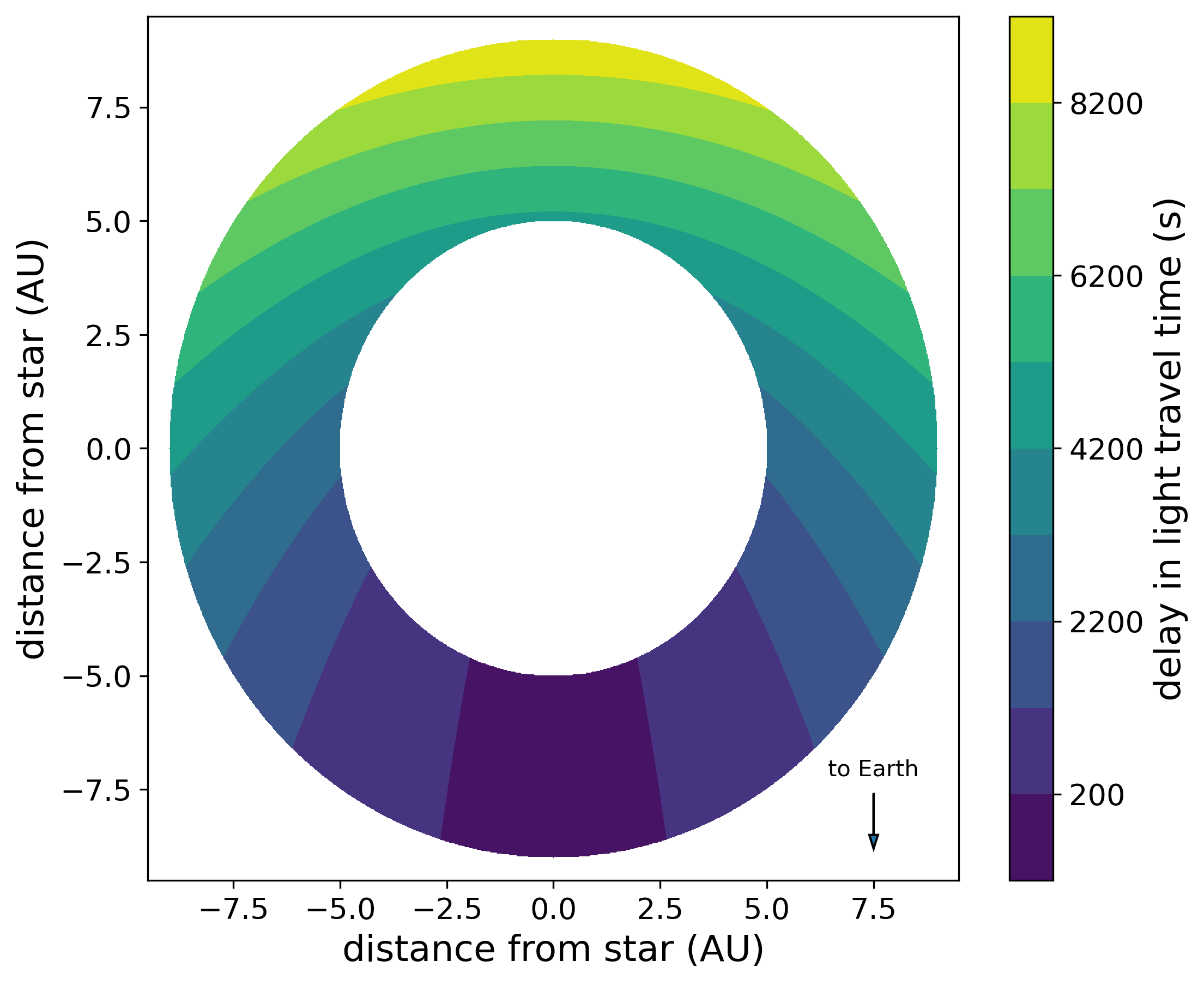}
\caption{The delay in light travel time, assuming the light went from the star to the spot in the disk before coming to us. Only parts of the disk in the darkest colored region, with delay times less than 200 s, are consistent with our measurements. If the emission can only come from the front part of the ring (to satisfy the time 
delay constraint), the line width of the line-of-sight emission would be smaller than the full 
velocity width of the ring.  \label{ltt}}
\end{figure}

    \begin{figure}
\centering
\includegraphics[width=3.2in]{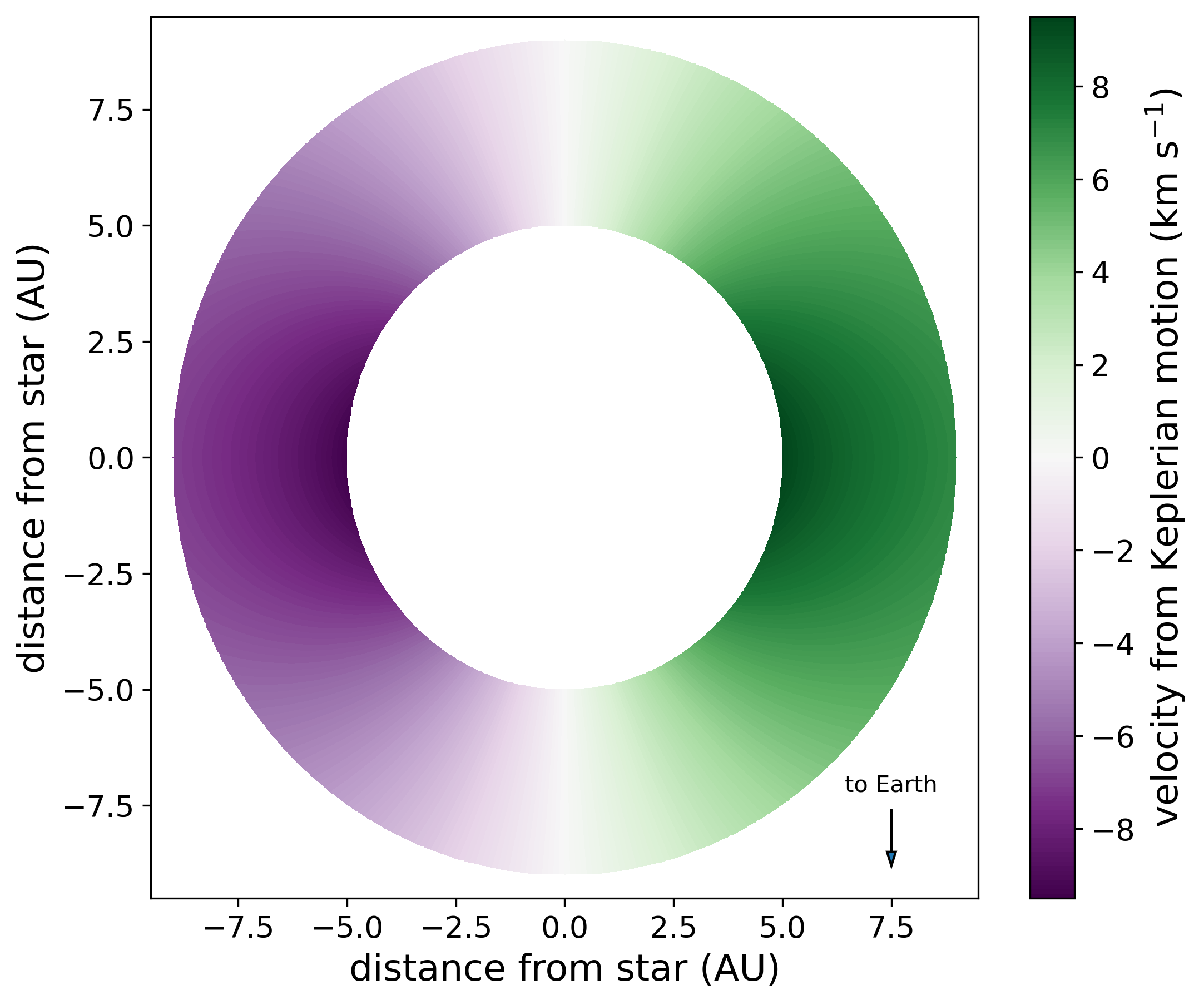}
\caption{The Keplerian velocity --- the dominant broadening component --- of gas in a disk between 5 and 9 AU, distances that would be consistent with the quiescent line profile.  For an edge-on disk, the broadness of the line profile is the result of adding the emission from all parts of the disk.    \label{disk_v}}
\end{figure}

  The response to the flare makes a disk origin even less probable.  The lines from [0,17] reach their peak in the flare up to 200 s after the \ion{C}{4} line flares (Figure \ref{lc_flare}).  This rules out most locations in the disk, as light from the flare would not have enough time to travel from the star to most locations in the disk, and then be reprocessed and redirected to us within 200 s.  In Figure \ref{ltt}, we show the regions of the disk where it would be physically possible to detect light with at most a 200 s delay.

  Additionally, the line profile from the flare is very similar to that during quiescence, with a FWHM of   19.0$\pm$1.2 km/s during the flare compared to 16.6$\pm$1.9 km/s in the 2020 quiescence observations.   For disk emission lines, the width of the line profile is due to emission at different velocities in different parts of the disk (Figure \ref{disk_v}).  If the source of the H$_2$ was a disk,  only a portion of the disk would be heated by the flare, and therefore the line profile would be significantly more narrow. Instead, the line profile from the flare has a very similar width to that from quiescence, as shown in Figure \ref{lsd}.

\subsubsection{Search for Absorption from the Disk in \ion{C}{4}}
  
If the  H$_2$ that is being pumped by \ion{C}{4} during the flare is at a few AU, we should be able to detect absorption in \ion{C}{4} if the gas is between us and the flare.  The scale height, $h$, of gas in a disk is approximately:

\begin{equation}\label{scalehight}
    h=c_s\sqrt{\frac{2r^3}{GM_*}}
\end{equation}
where $c_s$ is the sound speed \citep{HartmannAccretionProcessesStar2008}. Given a gas temperature, $T$, of 1500 K, and assuming all the gas is in H$_2$, so that the mean molecular weight $\mu$ is 2.016\footnote{https://pubchem.ncbi.nlm.nih.gov/compound/Hydrogen},  we can calculate the sound speed using $c_s=\sqrt{\frac{k_BT}{\mu m_H}}$=2.3 km/s, where $k_B$ is Boltzmann's constant and $m_H$ is the mass of hydrogen.  At even 3 AU, equation \ref{scalehight} gives a scale height of $\sim$175 R$_\odot$, increasing to $\sim$375 R$_\odot$ at 5 AU.   Given that the disk is almost edge on, with an inclination of at least 88 degrees \citep{DaleyMassStirringBodies2019}, the disk need only have a scale height of 22 R$_\odot$  at 3 AU or 38 R$_\odot$ at 5 AU to obscure the star, far less than the scale heights we calculate.  
 
\begin{figure*}[ht!]
\centering
\includegraphics[width=0.98\textwidth]{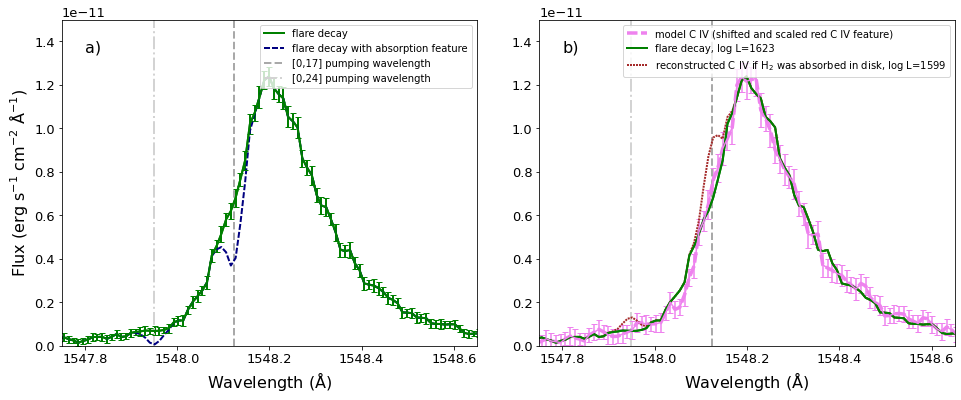}
\caption{(Left) The observed \ion{C}{4} profile (solid green line) during the flare shows no absorption features at the pumping wavelengths, which are marked with vertical, dashed, gray lines like we would expect (blue dashed line) if the H$_2$ was between the star and us. (Right) The observed flux matches  better with the modeled \ion{C}{4} profile (thick pink dashed line) than the reconstructed \ion{C}{4} profile (brown dotted line) matches the modeled profile. The reconstructed profile is calculated by adding on the potential absorption from the disk to the observed profile to recreate the intrinsic profile. Based on their corresponding likelihoods, the observed flux is significantly better match for the modeled \ion{C}{4} profile, as the reconstructed \ion{C}{4} can be clearly ruled out, with a p-value of $<$1$\times$10$^{-10}$ in comparison to the observed flux being the intrinsic profile.   \label{civ_absorption}}
\end{figure*}

We modeled the absorption we would expect to see from the disk at the pumping wavelengths with a Gaussian. The total flux absorbed out of the \ion{C}{4} line is measured using the observed fluxes in the pumped lines and their respective branching ratios.  When estimating the flux absorbed out of the \ion{C}{4} line in this way, we add back in the predicted flux to the pumping transition itself as the branching ratios predict that some fraction of the absorbed flux will be re-emitted at this same wavelength.  Thus, we estimate the actual absorption that would be seen at Earth.     For [0,17] we used the average absorbed flux based on the six strongest features; for [0,24] we took the average of both detected features.   The widths of the absorption features are those from thermal broadening at 1500 K plus the additional width from the line spread function  added in quadrature, while the total flux is the one calculated as described above. We then subtracted this absorption feature from the observed flux (Figure \ref{civ_absorption}a, solid green line) to see if such absorption would be detectable -- and it clearly is, as shown by the blue dashed line in the plot.  This calculation assumes the gas is distributed spherically; if the gas were in a column or a flat disk, the predicted absorption would be deeper.

Furthermore, if the observed flux was the result of H$_2$ absorption then adding that emission back on should result in the intrinsic \ion{C}{4} profile.  We modeled the true \ion{C}{4}  profile  (the dashed pink line in Figure \ref{civ_absorption}b) by using the observed flux from the red component of the \ion{C}{4} emission, centered at 1550.94 \AA. We Doppler shifted that observed red \ion{C}{4} profile and scaled it by a constant to match the blue component, creating a ``modeled'' \ion{C}{4} profile. The observed flux is a very good match for our modeled \ion{C}{4}, while the reconstructed \ion{C}{4} profile (the brown dotted line, calculated by adding on the potential absorption from the disk to the observed profile to recreate the corresponding intrinsic profile) is a poor match. We also calculated the likelihood for both the reconstructed \ion{C}{4} profile and the observed \ion{C}{4} profile in a similar manner to that of Equation \ref{modfit_ll}:
  \begin{equation}
       \log(L) =-\frac{1}{2}\sum_i\left[ \frac{(F_{mod,i}-F_{meas,i})^2}{\sigma_i^2} +\log(2\pi \sigma_i^2)\right]
\end{equation}
where in this case, the ``measured'' flux, $F_{meas,i}$, is the flux at each point from the shifted and scaled \ion{C}{4} profile that we assume to be the shape of the intrinsic profile, while the modeled flux is either the observed \ion{C}{4} profile flux or the reconstructed  \ion{C}{4} profile.  Based on their corresponding likelihoods, the observed flux is significantly better match for the intrinsic \ion{C}{4} profile, as the reconstructed \ion{C}{4} can be clearly ruled out, with a p-value of $<$1$\times$10$^{-10}$ in comparison to the observed flux being the intrinsic profile. Thus, we conclude that the  absorption in the \ion{C}{4} profile at the pumping wavelengths  that we would expect to see if the H$_2$ were between us and the \ion{C}{4} is not present.  Either the scale height we calculated is too large by an order of magnitude or the H$_2$ is not in the disk.

    \subsection{A Planet}
  While H$_2$ has not been conclusively detected in an exoplanet \citep[e.g][]{FranceSearchingFarultravioletAuroral2010, KruczekH2FluorescenceDwarf2017}, all models indicate that it should be prominent in gas giant exoplanets, similar to the gas giant planets in our Solar System \citep[e.g.,][]{SudarskyTheoreticalSpectraAtmospheres2003,YelleAeronomyextrasolargiant2004}.  This is especially true for young planets which have not yet undergone significant atmospheric escape. Since the effective temperatures of the planets orbiting AU Mic have not been measured, we cannot rule either planet out as sources of the H$_2$ based on their temperature. 

  \begin{figure}[h!]
\centering
\includegraphics[width=3.3in]{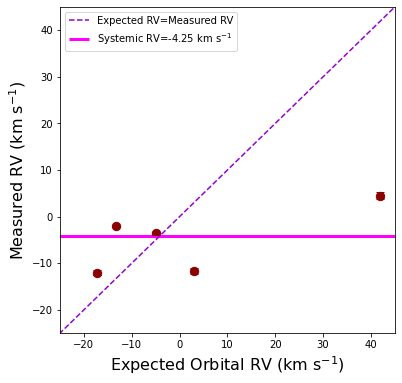}
\caption{The velocities we measured  from the 2020 versus what we would expect if the H$_2$ was in AU Mic b based on the orbit from \citet{Plavchanplanetdebrisdisk2020}.  We do not see the signal follow the planet's orbit --- instead it scatters around the systemic velocity of the star.  Uncertainties are plotted, but are generally smaller than the marker.  \label{rvs}}
\end{figure}

  However, in the 2020 observations, which span 23 hours (after discarding the observation with the flare), AU Mic b's projected velocity changes from -17 km/s to 42 km/s based on the orbital parameters from \citet{MartioliNewconstraintsplanetary2021}.  This is not reflected in the CCF.  In Figure \ref{rvs}, we plot the measured CCF velocity centers as a function of time, compared with the predicted velocity of the planet.  The velocities of the CCF are consistent with the systemic velocity --- which would be the expected central velocity for both the star and the disk --- but not AU Mic b's velocity.   AU Mic c can be ruled out for the same reason, as its expected velocities during these observations are all more than 30 km/s from the systemic velocity. Thus the H$_2$ line emission is not from either Au Mic b or c.  


  \subsection{The Star}\label{subsec_star}
\begin{figure*}
\centering
\includegraphics[width=0.98\textwidth]{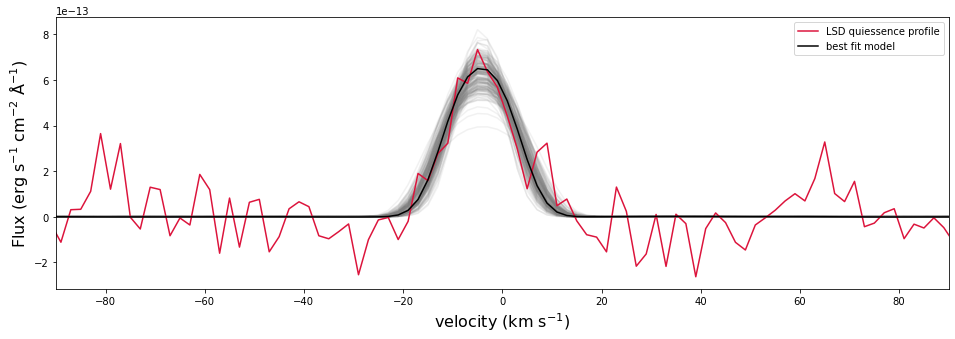}
\caption{The LSD profile plotted with the best fit from our MCMC simulation.  Also plotted are 200 samples to show the distribution of potential solutions. \label{mcmc_fit}}
\end{figure*}

\begin{figure}
\centering
\includegraphics[width=3.3in]{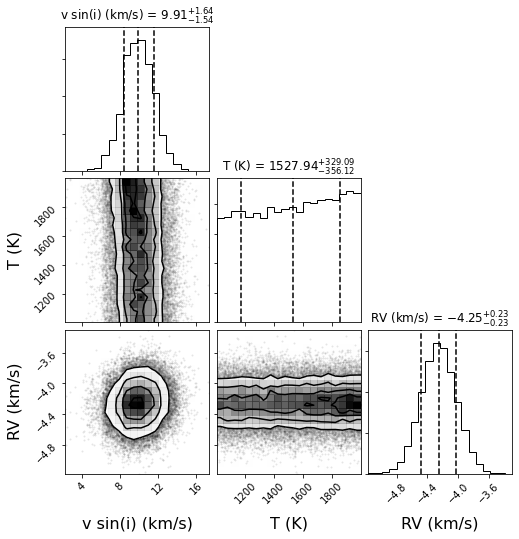}
\caption{Resulting distributions from the MCMC simulation. The median value is marked, as are the 16th and 84th percentiles, representing the 1$\sigma$ uncertainties. Our  value for $v\sin i$ is consistent with that measured from photospheric lines \citep{Plavchanplanetdebrisdisk2020}. \label{mcmc_corner}}
\end{figure}

By process of elimination, the most likely source for the observed H$_2$ is  the star, consistent with the inferred location of H$_2$ in older M dwarfs \citep{KruczekH2FluorescenceDwarf2017}. To check whether the star is a possible source, we fit  the LSD line profile from Figure \ref{profile_1998} with a simple model of an emission line arising on the stellar surface that includes  thermal broadening, rotational broadening, and instrumental broadening.  Our best fit from using a Markov Chain Monte Carlo (MCMC) algorithm is shown in Figure  \ref{mcmc_fit}. The best fit parameters, along with their corresponding priors are summarized in Table \ref{fitparam}; the full distribution from the MCMC for each parameter is shown in Figure \ref{mcmc_corner}. Based on this simulation, we estimate 9.91$^{+1.64}_{-1.54}$ km/s for the $v\sin i$, which is consistent with   8.7$\pm$2.0 km/s (Table \ref{params}) found by \citet{Plavchanplanetdebrisdisk2020} using photospheric lines.

 \begin{deluxetable}{lrrl} 
 \tabletypesize{\normalsize}
 \tablewidth{0pt}
 \tablecaption{Model Parameters}\label{fitparam} 
 \tablehead{
 \colhead{Parameter} & \colhead{    \hspace{.25cm}   } & \colhead{Best Fit Value} & \colhead{Priors}
  }
 \startdata
RV (km/s) & & -4.25$\pm$0.23 & $\mathcal{N}$(-4.25,0.24) \\
$v\sin i$ (km/s) & & 9.91$^{+1.64}_{-1.54}$ & $\mathcal{U}$(2,22)\\
$\rm{T_{ex}}$ & & 1528$^{+329}_{-356}$ & $\mathcal{U}$(1000,2000)\\
\enddata
\tablecomments{1$\sigma$ uncertainties are calculated based on the values at the 16th and 84th percentiles, as shown in Figure \ref{mcmc_corner}.}
 \end{deluxetable}

The profile during the flare is slightly broader and more blue shifted than the quiescent profile (Figure \ref{lsd}, Table \ref{profile_pars}).  We attribute this to the possible detection of mass motion of H$_2$ during the flare, similar to mass motion detected during solar flares \citep[e.g.][]{OhyamaPreflareHeatingMass1997}, as we cannot produce such a profile with the expected emission from H$_2$.

Based on expected temperatures, H$_2$ can exist in both the photospheres and starspots of M dwarfs.    AU Mic is not the only star with this cold H$_2$.    \citet{FranceHighenergyRadiationEnvironment2020} detected cold H$_2$ (which also brightened during a flare) in Barnard's Star, a $\sim$10 Gyr M3.5 dwarf. Given its age, Barnard's Star is very unlikely to have detectable quantities of residual circumstellar gas; thus, the H$_2$ must originate from the star itself.  

However, the measured temperature of the H$_2$ ($<$2000 K) is inconsistent with that of AU Mic's photosphere  \citep[$\rm{T_{eff}}$=3642$\pm$22 K;][]{PecautIntrinsicColorsTemperatures2013}.  So where in the star is it coming from?  We consider two possibilities: starspots and the temperature minimum in a cold layer between the photosphere and the chromosphere. 
  
  There are two main ways to estimate the starspot temperatures for AU Mic: by models or by observations.  There have been models of star spot temperatures for main sequence M0 dwarfs, which estimate temperatures of $\gtrsim$3000 K \citep{Panja3DRadiativeMHD2020}.  However, since AU Mic is more active than typical main sequence stars, with a radius that is 50\% larger than it will have once it reaches the main sequence \citep{BaraffeNewevolutionarymodels2015}, it is possible that these differences would also cause a change in starspot temperatures.  From an observational perspective, starspot temperatures measured specifically for AU Mic by  using photometric \citep{RodonoRotationalmodulationflares1986} or spectroscopic \citep{AframComplexitymagneticfields2019}) data  are far warmer than the H$_2$ temperature we measure, as are starspot temperature measurements for other young dwarfs \citep{Gully-SantiagoPlacingSpottedTauri2017}.  However, there is indirect evidence that starspots could be colder than these estimates.  In the Sun, which has a $\rm{T_{eff}}$ 2000 K warmer than AU Mic's, molecules in the sunspots have rotational and vibrational temperatures less than 2000 K \citep[e.g][]{MulchaeyRotationalTemperatureFeH1989, Sriramachandraneffectivetemperaturelarge2011}, indicating starspots might also get cooler than models predict.  Furthermore, starspots need not all be the same temperature \citep[e.g.][]{KoppRelationMagneticField1992}.  If some starspots are at temperatures corresponding to the ones measured by \citet{AframComplexitymagneticfields2019}, we do not think this completely excludes the possibility of colder starspots. Plus, like the photosphere, an individual starspot has layers of different temperatures.    Therefore, we cannot rule out starspots across the star as a potential source for the H$_2$.    

  The other possibility is a layer of colder gas at the top of the photosphere near the temperature minimum.  In the Sun, this layer is called the CO-mosphere \citep{AyresDoesSunHave2002}, because it is cool enough (T$\sim$3500 K) for CO to form compared to the Sun's effective temperature of 5800 K \citep{NoyesSpectraCOFundamental1972, AyresFourierTransformSpectrometer1981, WiedemannCarbonMonoxideFundamental1994}.  In the case of AU Mic, the photosphere is cold enough for CO to form anywhere, so the name of the layer would be different, but the basic structure of a colder layer could easily hold.  Cold layers occur in M giants, producing H$_2$O that cannot exist in those stars' deeper photospheres \citep{SloanInfraredSpectralProperties2015}, but similar structures have not been detected in M dwarfs.  Current models indicate that there should be a cold layer between the photosphere and the chromosphere, but again, the modeled gas does not get cold enough, as it is predicted to only reach $\sim$2500 K \citep{FontenlaSemiempiricalModelingPhotosphere2016}.  Still, as the physics of the temperature structure in the top of the photosphere for M dwarfs is undoubtedly complicated and has not been extensively studied, we think it is possible that current models do not capture all the physics of such star's atmospheric structure.
  
  There is a third possibility.  Our temperature calculations assume that H$_2$ is populated thermally, because models of disk heating are consistent with thermal heating \citep{AdamkovicsFUVIrradiatedDisk2016}.  While there is no evidence of this, we cannot rule out some degree of non-thermal population that would skew our temperature measurement.  However, in the case of Jupiter, temperature estimates from line ratios are warmer than the kinetic temperatures estimates \citep{BarthelemyH2vibrationaltemperatures2005}.  Jupiter is obviously much colder than the temperatures we measured, but if the same holds true for AU Mic, the temperature of the H$_2$ is even less than our 1000-2000 K estimate and would not explain the temperature discrepancy.

  \section{Conclusions}
  In this paper, we report on our analysis of H$_2$ in AU Mic including: 
  \begin{itemize}
      \item detecting H$_2$ in AU Mic from HST-STIS FUV spectra both during quiescence and a flare
      \item measuring the temperature of the H$_2$ during quiescence at $>$1000 K
      \item characterizing the response of the H$_2$ to a stellar flare, showing non-thermal emission pumped by \ion{C}{4}
      \item ruling out a foreground/background source, the circumstellar disk or a planet as the source of this H$_2$ 
  \end{itemize}
Based on the line profile and the response to a stellar flare, we conclude that the only possible source of the H$_2$ is in the star itself.  However, the temperatures we measure indicate that this gas is too cold to be from the star based on current models of M dwarfs and their spots.
  
  This detection obviously presents a mystery. Current models cannot account for this H$_2$ --- but it seems clear that the H$_2$ emission must be produced by the star. 

\acknowledgements
We thank the anonymous referee for their helpful comments.  Based on observations with the NASA/ESA Hubble Space Telescope obtained at the Space Telescope Science Institute, which is operated by the Association of Universities for Research in Astronomy, Incorporated, under NASA contract NAS5-26555. Support for program number (GO-15310) was provided through a grant from the STScI under NASA contract NAS5-26555. GJH is supported by by general grant 12173003 awarded by the National Science Foundation of China.   This research has made use of the VizieR catalogue access tool, CDS, Strasbourg, France. The original description of the VizieR service was published by  \citep{WengerSIMBADastronomicaldatabase2000}.  This research has made use of the SIMBAD database,operated at CDS, Strasbourg, France. 
  
\facility{HST (STIS)}

\software{SpecTres \citep{CarnallSpectResFastSpectral2017}, NumPy \citep{oliphant2006guide, van2011numpy},  Pandas \citep{reback2020pandas},  Matplotlib \citep{Hunter:2007}, LMFIT \citep{NewvilleLMFITNonLinearLeastSquare2014}, emcee \citep{Foreman-MackeyemceeMCMCHammer2013}}

\restartappendixnumbering

\appendix
\section{\texorpdfstring{Individual H$_2$ Features from 2020 Flare}{Individual H2 Features from 2020 Flare}}

\begin{figure*}
\centering
\includegraphics[width=0.83\textwidth]{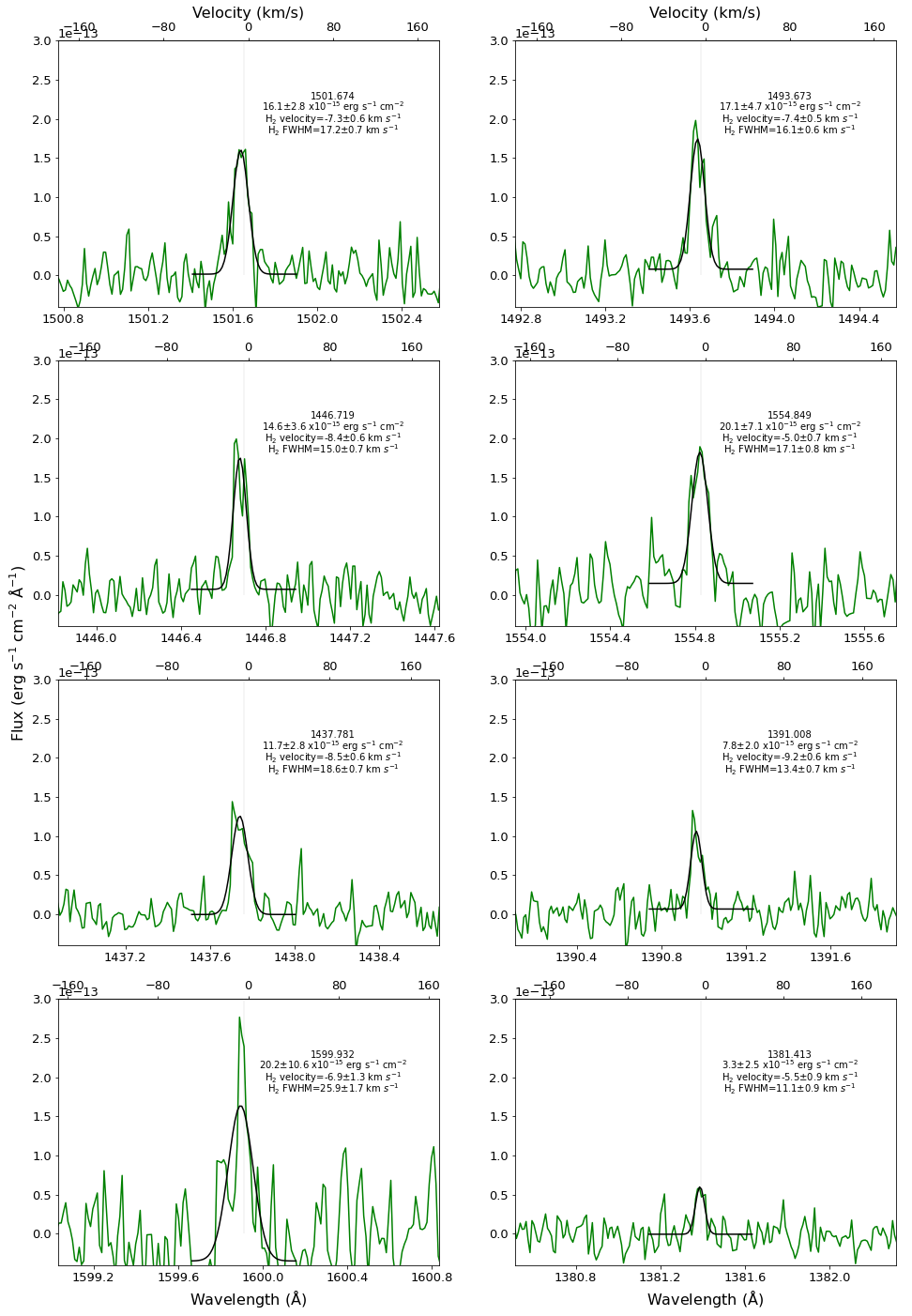}
\caption{Gaussian fits to H$_2$ profiles for features in the [0,17] progression from observation OE4H01010 acquired on 2020 July 2.   \label{h2_017}}
\end{figure*}

\begin{figure*}
\centering
\includegraphics[width=0.98\textwidth]{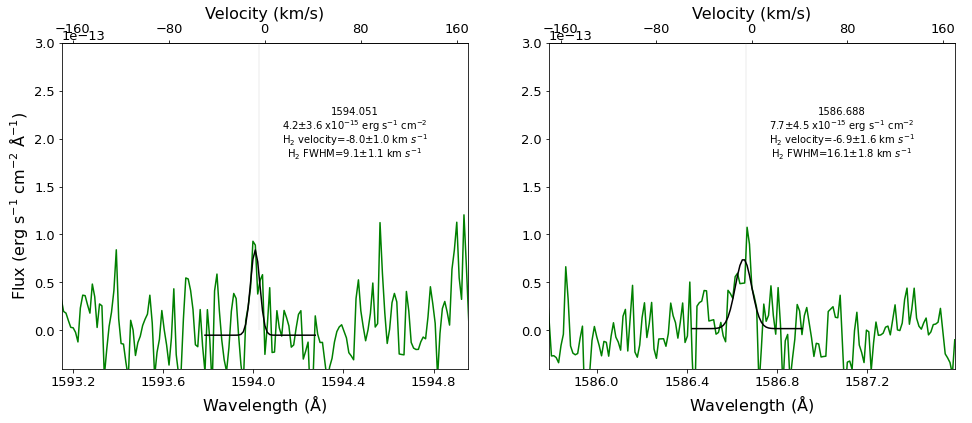}
\caption{Gaussian fits to H$_2$ profiles for features in the [0,24] progression from observation OE4H01010 acquired on 2020 July 2.   \label{h2_024}}
\end{figure*}

The H$_2$ features  detectable by eye during the 2020 flare are from the [0,17] (Figure \ref{h2_017}) and [0,24] (Figure \ref{h2_024}) progressions.

\bibliography{h2indisks_url}{}
\bibliographystyle{aasjournal}



\end{document}